# The Shrinking Sweet Spot: How Algorithms, Institutions, and Social Priors Shape Musical Ecosystems

Fabio Lokwani Di Matteo and Pier Luigi Sacco*

*Department of Neuroscience, Imaging and Clinical Sciences,*

*Università G. d'Annunzio, Chieti-Pescara, Italy*

* Corresponding author: pierluigi.sacco@unich.it, pierluigi_sacco@fas.harvard.it

**Abstract**

Why do some national music markets sustain a rich musical diversity whereas others converge on mostly formulaic output? The existing models of cultural consumption (superstar economics, rational addiction, Bayesian social learning) each capture part of the answer, but none can explain how exposure, social influence, institutional gatekeeping, and algorithmic curation interact to shape what listeners come to prefer. We address this gap by modeling musical taste as a learning process rather than a fixed parameter: a listener's evaluative disposition evolves with each encounter, shaped by the balance between the comfort of the familiar and the reward of the new. Drawing on the active inference framework from cognitive science, we formalize this as a sequential choice model in which preferences, information, and the consumption environment co-evolve, and show how the framework nests and extends key mechanisms from the three canonical economic models. An agent-based simulation generates four predictions: algorithmic curation suppresses consumption diversity beyond a sharp nonlinear threshold; institutional structure determines winner-take-all intensity through confirmatory cross-system contrasts; cultural capital buffers listeners against homogenization; and high-curation, high-conformity systems collapse supply-side dispersion relative to pluralistic ecosystems. We test the framework against four national music ecosystems (Italy's Festival di Sanremo, Brazil, South Korea, and the United Kingdom), identifying structural determinants of ecosystem vitality on both the supply and demand sides. The welfare implications are direct: because listeners' preferences adapt to impoverished environments through the very learning mechanisms the model describes, revealed preference analysis cannot reliably evaluate the outcomes of cultural markets. We ground the case for intervention in Sen's capability approach, Elster's adaptive preferences, and positive externalities of diversity.

# 1. Introduction

What makes a pop song actually popular? The question appears deceptively simple, yet it has resisted rigorous explanation across disciplines. Music industry professionals routinely fail to predict hits (Elberse, 2013). Algorithms trained on vast catalogues achieve only modest predictive accuracy (Pachet & Roy, 2008). And listeners themselves, when asked to explain their preferences, tend to invoke the language of intrinsic quality (a song is 'catchy,' 'moving,' or simply 'good') as though popularity were a natural consequence of inherent musical properties.

The stakes extend well beyond aesthetics. The global recorded music industry generated over $28 billion in revenues in 2023 (IFPI, 2024). National music scenes serve as vectors of soft power, identity formation, and cultural exports. Yet the mechanisms through which certain songs achieve commercial and cultural traction remain poorly theorized within cultural economics. The dominant frameworks such as Rosen's (1981) superstar economics, Adler's (1985; 2006) consumption capital theory, Becker and Murphy's (1988) rational addiction model, and Bayesian social learning (Banerjee, 1992; Bikhchandani, Hirshleifer & Welch, 1992) each capture important mechanisms but remain siloed, unable to explain how cognitive, social, institutional, and algorithmic processes interact to produce the observed dynamics of cultural markets. The problem is conceptual, not only empirical. Cultural economics has long acknowledged that preferences for cultural goods are endogenous, shaped by the very market processes they are supposed to evaluate, yet the field lacks a formal apparatus that makes this endogeneity tractable rather than simply noting it as a complication.

We propose that the active inference framework (Friston, 2010; Parr, Pezzulo & Friston, 2022) provides the missing integrative architecture. The core idea is intuitive. Listeners are not passive consumers matching songs against a stable preference ordering; they are learners whose evaluative dispositions are continuously reshaped by what they hear. Each listening experience updates an internal model of what music sounds like, what to expect, and what is rewarding. Familiarity breeds fluency: a song heard several times becomes easier to process, and that fluency is experienced as pleasure. But familiarity also breeds satiation: once a song holds no surprises, the reward fades. The sweet spot, the region where a song is familiar enough to be engaging yet novel enough to be interesting, is not a fixed property of the song; it is a moving target that depends on the listener's accumulated experience. Active inference formalizes this dynamic by recasting 'taste' as the current state of a predictive model that is updated with every encounter. In this framework, the pleasure of music arises not from matching a pre-existing preference but from the interplay between expectation and surprise: a song rewards the listener when it is partly predictable (confirming what the model has learned) and partly surprising (providing new information that updates the model further). The economic parallel is the exploration-exploitation tradeoff: listeners balance the reliable reward of the familiar against the potential discovery value of the new.

This formulation does real analytical work. It explains why the same person can find a song boring on first hearing, captivating after several plays, and tedious after dozens, a trajectory that no fixed-preference model can generate. It explains why social influence amplifies success: popular songs receive more exposure, which shifts them into the sweet spot for more listeners, which generates more plays, in a feedback loop that is only loosely coupled to any intrinsic property of the music. It explains why algorithmic curation, which selects for songs that are reliably pleasant to the largest number of listeners, systematically narrows the musical landscape: algorithms optimize for the familiar end of the sweet spot, eroding the novelty that sustains long-run engagement. And it explains why certain national music scenes undergo progressive impoverishment. When social influence, institutional gatekeeping, and algorithmic curation interact in feedback loops that shrink the range of music that listeners encounter, the result is convergence on formulaic output that maximizes short-run comfort at the cost of what the framework terms 'epistemic value': the discovery reward that comes from encountering something genuinely new, corresponding in economic terms to the option value of cultural exploration. We illustrate this through the case of Italy's Festival di Sanremo, contrasting it with structurally healthier ecosystems in Brazil, South Korea, and the United Kingdom.

The paper is organized as follows. Section 2 reviews the empirical evidence on taste construction. Section 3 introduces the active inference framework and explicitly compares it with existing economic models. Section 4 develops the integrative account of how pop songs become popular. Section 5 provides the formal specification, multi-agent extension, and computational simulation. Section 6 applies this to national music scenes with a cross-national comparison. Section 7 grounds policy recommendations in welfare theory, and Section 8 concludes.

## 2. The Construction of Musical Taste: Converging Evidence

Before introducing the formal framework, we review several bodies of evidence that collectively show that musical taste is not a stable individual attribute but a dynamically constructed property of the interaction between listeners, social environments, institutions, and technologies. Each body of evidence is well established individually; the contribution of this section is to show that they converge on a single implication that existing economic models have not fully absorbed: taste is a process, not a parameter.

The most striking evidence for the constructed nature of musical taste comes from Salganik, Dodds, and Watts's (2006) MusicLab experiments. Over 14,000 participants were recruited to rate and download songs by unknown bands in an online market. Participants were randomly assigned to one of two conditions: an independent condition (no information about others' choices) or one of eight social influence 'worlds' (where download counts were visible). The results were remarkable. In the independent condition, song quality, as revealed by average ratings, was a modest predictor of market share. But in the social influence conditions, the eight parallel worlds produced dramatically different popularity rankings from identical song pools. Songs that achieved massive popularity in one world languished in obscurity in another. Quality imposed only weak constraints: for the vast majority of songs, virtually any market outcome was possible.

A follow-up experiment (Salganik & Watts, 2008) sharpened the finding. When the experimenters deliberately inverted the initial popularity rankings (placing the least popular songs at the top of the chart), these artificially promoted songs maintained much of their induced advantage, though the very best songs eventually recovered some market share. The implication is that in systems with social influence, the relationship between intrinsic quality and commercial success is mediated by path-dependent cumulative advantage dynamics (Watts, 2011). Success breeds success through a mechanism that is only loosely coupled to any property of the product itself. For economic theory, this finding poses a fundamental challenge: if the same product can achieve radically different market outcomes depending on the sequence of early adopters, then standard welfare analysis based on revealed preference becomes deeply problematic. Consumers are not revealing pre-existing preferences; they are constructing preferences through a socially mediated process.

A second mechanism operates at the individual cognitive level. Bornstein's (1989) meta-analysis of 208 studies established that repeated exposure to a stimulus increases preference for it, with a robust effect size of r = 0.26. For music, the effect operates even under subliminal conditions (Monahan, Murphy & Zajonc, 2000) and follows the inverted-U curve first described by Berlyne (1971): preference increases with moderate repetition but decreases with overexposure. Huron (2006) provided the critical reinterpretation: the mere exposure effect is a *prediction effect*. Repeated exposure enables the listener to build an increasingly accurate predictive model of the stimulus, and the resulting processing fluency generates positive affect. This reinterpretation transforms the mere exposure effect from a psychological curiosity into a window onto the computational processes underlying taste formation.

The economic significance of this mechanism is substantial. It means that preference is not independent of the consumption history that standard demand theory treats as given. The listener who has heard a song ten times has a different cognitive relationship to it, not simply more information. Her predictive model has been updated, her processing fluency has increased, and her affective response has changed accordingly. This is not the accumulation of consumption capital in Becker and Murphy's (1988) sense, where the utility function remains fixed while consumption capital grows, but a transformation of the evaluative apparatus itself. The inverted-U dynamics also imply that there exists an optimal exposure level beyond which further repetition produces satiation and declining preference, a feature that has direct consequences for the design of recommendation algorithms and that we formalize in Section 5.1 through the Dirichlet concentration parameter mechanism.

The third mechanism is technological. Anderson, Maystre, Anderson, Mehrotra, and Lalmas (2020), analyzing tens of millions of Spotify listening sessions, found that algorithmically-driven listening is consistently less diverse than user-driven listening. When listeners select music themselves, they explore a wider range of artists and genres than when they follow algorithmic recommendations. This provides large-scale evidence that recommendation systems, optimized for engagement metrics, systematically narrow the listener's musical environment.

However, the relationship between algorithmic curation and diversity is not monolithic. Platform architectures embody different optimization objectives that produce different ecological effects. TikTok's algorithm, designed to maximize content virality rather than personalized comfort, has been credited with surfacing unexpected regional content, enabling Brazilian funk carioca and pisadinha to reach national audiences and propelling K-pop acts to global visibility through short-form video engagement. Spotify's personalization engine, by contrast, tends toward what we might call comfort zone maintenance: algorithmically generated playlists that match the listener's established taste profile with increasing precision (Prey, 2020; Bonini & Magaudda, 2024). YouTube's recommendation system optimizes for session length, which may favor genre consistency within listening episodes but allow greater diversity across sessions (Eriksson, Fleischer, Johansson, Snickars & Vonderau, 2019). These platform-specific differences are theoretically important: they show that algorithmic curation is not a single force but a family of mechanisms whose effects on musical ecosystems depend on the specific computational objectives and interface architectures through which they operate. Any adequate theory must accommodate this heterogeneity.

Yet another mechanism is highlighted by Bourdieu's (1984) analysis in *Distinction*, showing that musical preferences are not merely expressions of individual personality but systematic markers of social position, mediated by the *habitus*, a system of durable dispositions acquired through socialization that generates practices and perceptions without presupposing a conscious strategy. Subsequent research refined this picture in important ways. Peterson and

Simkus (1992) introduced the 'cultural omnivore' thesis, showing that high-status individuals in contemporary societies tend not toward exclusive highbrow consumption (as Bourdieu's original formulation suggested) but toward broader taste repertoires spanning multiple genres and traditions. Life-course studies (van Hek & Kraaykamp, 2015) established that education predicts taste breadth more strongly than income, suggesting that the relevant resource is cognitive rather than financial.

We adopt Bourdieu's insights selectively, and this selectivity requires explicit acknowledgment. The active inference framework captures the generative and dispositional aspects of habitus well: the habitus as a set of internalized schemes that generate context-appropriate responses without conscious deliberation maps naturally onto a hierarchical generative model. However, Bourdieu's framework is fundamentally *relational*: taste operates as a mechanism of social distinction that simultaneously classifies cultural objects and the agents who consume them within a field of power relations. This relational dimension (the way that preferring jazz over pop simultaneously positions the listener within a social hierarchy) is not captured by the active inference formalism, which models individual agents' generative models without embedding them in a field-theoretic structure. Similarly, Hennion's (2007) complementary insight that taste is a distributed, relational practice involving material, bodily, and collective mediations, also requires theoretical machinery beyond our scope. We acknowledge these limitations rather than claiming a full integration.

We do note, however, one structural parallel that partially bridges the gap: the alignment prior in multi-agent active inference (Constant et al., 2019), the tendency for agents to seek evidence that their mental states match those of conspecifics, is formally analogous to Bourdieu's insight that taste operates as social recognition. Both frameworks predict that individuals will gravitate toward cultural consumption patterns that signal membership in valued social groups. The mechanisms differ (Bayesian updating on social evidence versus field-mediated distinction), but the behavioral predictions converge.

The construction mechanisms reviewed above predict a tendency toward homogenization under certain institutional conditions. Serrà, Corral, Boguñá, Haro, and Arcos (2012) provided a direct empirical confirmation. Analyzing 464,411 recordings spanning 1955–2010 using signal processing techniques applied to the Million Song Dataset, they found three robust trends: progressive restriction of pitch transitions (the range of harmonic intervals used in popular music has narrowed), homogenization of the timbral palette (recordings have converged toward a smaller set of instrumental textures and production techniques), and steady increase in loudness (the so-called 'loudness war'). These are objective, measurable properties of the musical signal itself, independent of listener reports or commercial data. They confirm that the social, psychological, and technological mechanisms described above have material consequences for the characteristics of the music that is produced and consumed.

The convergence of these five evidence streams (social influence, processing fluency, algorithmic curation, cultural capital, and measurable homogenization) points toward a clear theoretical need: a unified framework that can accommodate all five mechanisms within a single formal architecture, explaining how they interact to produce the observed dynamics of cultural markets. The active inference framework, to which we now turn, provides precisely this.

## 3. Active Inference, Musical Pleasure, and Existing Economic Frameworks

### 3.1 The Free Energy Principle and Music as Hierarchical Prediction

The motivation for importing a cognitive framework into cultural economics is not interdisciplinary fashion but analytical necessity. The endogeneity of preferences that complicates welfare analysis in cultural markets is not an abstract theoretical possibility; it reflects specific, empirically documented mechanisms by which exposure, prediction, and reward interact in the brain. Understanding these mechanisms constrains which economic models of preference formation are empirically admissible, ruling out, for example, models that treat preference change as a simple stock variable and admitting instead models where the evaluative apparatus itself transforms through experience.

Active inference, a corollary of the free energy principle (Friston, 2010), describes how biological agents minimize surprise by continuously updating internal generative models. Music is a quintessentially temporal stimulus organized at multiple hierarchical levels: individual notes, melodic phrases, harmonic progressions, sections, and large-scale form (Friston & Friston, 2013). Van de Cruys, Frascaroli, and Friston (2024) extended this to aesthetic experience, proposing an 'epistemic arc,' a temporal trajectory from curiosity (high uncertainty, high expected information gain) through epistemic action (active exploration) to aha-experience (surprise resolution). This arc structure is critical for understanding how individual songs generate aesthetic reward, and, crucially for our argument, how impoverished musical ecosystems truncate or prevent the arc. When songs minimize prediction error from the outset, they provide no curiosity phase and therefore no epistemic action or resolution.

Neuroimaging evidence directly supports the computational account. Cheung, Harrison, Meyer, Pearce, Haynes, and Koelsch (2019), using a machine-learning model (IDyOM; Pearce, 2005) to quantify the uncertainty and surprise of 80,000 chords in US Billboard pop songs, found that pleasure varied nonlinearly as a function of the interaction between prospective uncertainty and retrospective surprise. Activity in the amygdala, hippocampus, and auditory cortex reflected this interaction, while the nucleus accumbens encoded uncertainty, consistent with the dopaminergic encoding of precision in the free energy framework. Salimpoor, Benovoy, Larcher, Dagher, and Zatorre (2011) showed that dopamine release in the striatum during musical chills correlates with peak emotional responses, confirming that the reward system is engaged by prediction-related processes in music. These findings provide neural evidence for a listener-relative 'sweet spot' of musical pleasure: maximum reward arises where the song is sufficiently familiar for model tracking but sufficiently novel for epistemic gain. This sweet spot depends on model precision and complexity; it is different for every listener and changes as the listener's model updates. The term 'epistemic value' here refers specifically to the expected information gain from model updating, not to epistemological evaluation in the philosophical sense.

### 3.2 What Active Inference Adds to Existing Economic Models

Before developing the formal model, we address the question of theoretical value added: what does the active inference framework contribute that is not already available in our existing toolkit? We argue that it provides three specific advances over current approaches to cultural consumption, and that these advances address gaps that the field has recognized but not yet resolved. Table 1 summarizes; we develop the argument below.

**Table 1: Comparison of Active Inference with Existing Economic Frameworks**

| Dimension | Rational Addiction (Becker & Murphy, 1988) | Bayesian Social Learning (Banerjee, 1992) | Superstar Economics (Rosen, 1981; Adler, 2006) | Active Inference (this paper) |
|---|---|---|---|---|
| **Preference structure** | Fixed utility function; consumption capital stock changes | Fixed prior beliefs updated by social observation | Exogenous preferences with imperfect substitution | Endogenous: generative model *is* the preference structure |
| **Novelty vs. familiarity** | No formal distinction | Not modeled | Not modeled | Decomposed into epistemic value (information gain) and pragmatic value (hedonic reward) |
| **Agent role** | Consumer with rational foresight | Passive observer of others' actions | Consumer selecting among imperfect substitutes | Active sampler selecting observations based on expected free energy |
| **Information environment** | Exogenous prices, income | Exogenous signal sequence | Exogenous quality distribution | Endogenous: niche construction reshapes the transition dynamics |
| **Social influence mechanism** | Peer effects on consumption capital | Information cascades from observed actions | Consumption capital spillovers | Precision-weighted social priors on prior distribution; alignment prior |
| **Population dynamics** | Steady-state addiction equilibria | Cascade/herd equilibria | Winner-take-all market shares | Convergent feedback loops with institutional modulation |
| **Neural substrates** | Not specified | Not specified | Not specified | Dopaminergic precision encoding; amygdala-hippocampal surprise processing |

### 3.2.1 Beyond Rational Addiction: Endogenous Preferences and Epistemic Value

Becker and Murphy's (1988) rational addiction model provides the closest existing analog to our exposure-dependent learning mechanism. In their framework, current utility depends on a stock of 'consumption capital' accumulated through past consumption, generating adjacent complementarity: past listening increases the marginal utility of current listening. This captures an important aspect of how musical taste develops over time. However, the rational addiction model is limited in three ways that active inference transcends.

First, rational addiction assumes *stable preferences*. The utility function is fixed; what changes is the consumption capital stock S, not the preference ordering itself. In Stigler and Becker's (1977) formulation, *de gustibus non est disputandum*: tastes do not change, only the household production function does. Active inference rejects this separation. The generative model *is* the preference structure: as the model updates through exposure and learning, what the listener finds rewarding changes endogenously. A listener who has spent a year immersed in jazz does not merely have a larger stock of jazz consumption capital; she has a *different generative model* that makes different predictions, finds different prediction errors rewarding, and consequently experiences different aesthetic pleasures. Preferences are not stable parameters acted upon by consumption capital; they are the dynamic output of the learning process itself.

Moreover, rational addiction has no concept of *epistemic value*. In Becker and Murphy's framework, the reward from listening is entirely hedonic: it enters through the utility function as a function of current consumption and accumulated capital. There is no formal distinction between the pleasure of hearing something familiar and the pleasure of learning something new. Active inference's decomposition of expected free energy into pragmatic value (hedonic reward from preferred observations) and epistemic value (information gain from model updating) captures a distinction that listeners readily recognize: the comfortable pleasure of a well-known song is qualitatively different from the thrilling pleasure of a novel chord progression that resolves unexpectedly. Cheung et al.'s (2019) neuroimaging data confirm that these are neurally dissociable processes. This distinction is not merely aesthetic; it has direct economic consequences, because epistemic value is what drives exploration of new music, whereas pragmatic value drives exploitation of the familiar. The balance between them determines the dynamics of musical ecosystems.

Finally, rational addiction cannot explain *why the same person might rationally choose both novelty-seeking and familiarity-seeking behavior* depending on context. In the Becker-Murphy framework, the consumption capital stock determines a single trajectory. Active inference predicts context-dependent switching between exploration and exploitation, governed by the relative precision of prior beliefs and the expected information gain from different policies. A listener may seek familiar music when cognitively depleted (low capacity for model updating) and novel music when alert and curious (high expected epistemic value), a pattern well-documented in the music psychology literature (Park, Thom, Mennicken, Cramer & Macy, 2019) but unaccounted for by rational addiction.

### 3.2.2 Beyond Bayesian Social Learning: Active Sampling and Niche Construction

The Bayesian social learning literature (Banerjee, 1992; Bikhchandani, Hirshleifer & Welch, 1992; Smith & Sørensen, 2000) provides well-established models of how rational agents update beliefs by observing others' actions. Information cascades arise when the accumulated weight of public information overwhelms private signals, leading agents to disregard their own preferences and follow the herd. The MusicLab results have been interpreted within this framework, and the interpretation captures important dynamics.

Active inference subsumes and extends this account in two ways. First, Bayesian social learning models treat the agent as a *passive observer* of others' actions. The agent updates beliefs based on information that arrives exogenously. Active inference, by contrast, models the agent as an *active sampler* who selects which observations to seek based on their expected epistemic and pragmatic value. A listener actively decides whether to listen to a popular song, and this decision is governed by the expected free energy of the 'listen to popular song' policy versus alternative policies. This is not a minor technical distinction: it determines whether the agent explores or exploits, and thus whether information cascades are self-reinforcing or self-limiting. An agent who is merely observing and updating will follow the cascade indefinitely; an agent who is actively sampling and evaluating epistemic return may break the cascade when the expected information gain from the popular song drops below that of an unexplored alternative.

Second, and more fundamentally, Bayesian social learning models treat the *information environment as exogenous*. Agents learn from a stream of observations whose statistical properties are independent of the agents' own learning behavior. Active inference, through the concept of niche construction (Constant, Ramstead, Veissière & Friston, 2019), recognizes that agents' actions reshape the environment from which future observations are drawn. In music markets, this is what algorithmic curation does: the listener's past choices alter the statistical properties of the musical environment that algorithms present. The transition dynamics in the POMDP model below are non-stationary and policy-dependent, a feature that standard Bayesian social learning models do not accommodate. We note that recent work on endogenous networks in social learning (Golub & Sadler, 2016) incorporates some agent influence over information structure; active inference goes further by making the statistical properties of the environment (not just the network topology) dependent on agents' actions. This is why the same song pool can produce radically different ecosystems depending on institutional structure: different agents not only have different priors; their collective learning behavior reshapes the environment itself.

### 3.2.3 Beyond Superstar Economics: Grounding Winner-Take-All Dynamics in Cognition

Rosen's (1981) superstar model explains extreme income inequality through the interaction of imperfect substitutability and joint consumption technology: when consumers prefer the best, and technology allows the best to serve many consumers at low marginal cost, market shares become highly concentrated. Adler (1985, 2006) added that consumption capital (knowing more about an artist makes it cheaper to enjoy their work) generates path-dependent dynamics that concentrate success independently of talent differentials.

Active inference does not replace these models but provides the *micro-cognitive foundations* they lack. Rosen and Adler explain *that* winner-take-all dynamics occur but not *why* particular winners emerge from specific environments. The active inference framework provides the cognitive mechanism: a song becomes a 'superstar' when it occupies the sweet spot of the population's collective generative model, simultaneously satisfying familiarity constraints (high pragmatic value), generating rewarding prediction error (high epistemic value), and benefiting from precision-weighted social priors (cumulative advantage). Crucially, this grounding explains something that superstar economics alone cannot: *why different institutional environments produce different superstar dynamics*. The same population of listeners can produce a healthy, diverse ecosystem or a convergent, impoverished one depending on the institutional structures that shape the distribution of generative sources, because institutions alter the transition dynamics (the rules governing how the available musical environment changes over time) and the prior distribution (the initial expectations listeners bring to the market), not just the marginal cost of distribution. This generates a specific prediction: the bandwidth of the superstar sweet spot varies with the variance of the population's preference distribution. Populations with narrow distributions (homogeneous taste cultures) produce steeper winner-take-all dynamics and fewer simultaneous superstars than populations with broad distributions. Our simulation analysis illustrates this prediction (Section 5.3, Figure 2B).

The relationship between the active inference framework and the three canonical models is not merely one of extension but of formal nesting. Each existing model can be recovered as a limiting case of the active inference approach by imposing specific constraints on its elements. *Rational addiction* emerges when the preference structure is held fixed (preferences do not change with exposure, only the consumption capital stock does), when there is no distinction between the reward of novelty and the reward of familiarity, and when the consumption environment is exogenous (not shaped by the agent's choices). Under these constraints, the model reduces to an agent accumulating exposure to a stimulus whose hedonic value depends on a stock variable, that is, the Becker–Murphy structure. *Bayesian social learning* emerges when the agent's role is restricted to passive observation (the agent observes others' choices and updates beliefs but does not actively select which stimuli to sample), when direct experience with songs does not update the agent's evaluative model, and when social signals dominate the decision to listen. Under these constraints, the agent follows the herd whenever the accumulated social evidence exceeds the private signal: the informational cascade of Banerjee (1992) and Bikhchandani et al. (1992). *Superstar economics* emerges when social influence is strong, when the population's preference distributions are homogeneous (all listeners have similar sweet spots), and when all listeners face the same choice set. Under these constraints, small quality differences interact with social amplification and shared access to produce the extreme concentration of market shares that Rosen (1981) and Adler (1985) describe. None of these limiting cases requires the full active inference machinery, which is why the individual models are analytically tractable. But none of them can generate the dynamics that arise when the constraints are relaxed simultaneously, i.e. when preferences, information, social signals, and the consumption environment co-evolve. That is the analytical space the active inference framework occupies.

## 4. How Pop Songs Become Popular: An Integrative Account

We now integrate the empirical evidence of Section 2 with the theoretical framework of Section 3 to explain how pop music popularity emerges from the interaction of four mechanisms, each formalizable within active inference.

In a social environment, others' choices function as precision-weighted priors. When a listener observes high stream counts, this constitutes a social observation that updates beliefs about the song's expected quality as a source of prediction error. The MusicLab results (Section 2.1) follow directly: in the independent condition, listeners sampled based on their own priors; in the social influence condition, download counts provided additional precision-weighted priors that dominated sampling policy. Because early downloads are partially random, different worlds developed different social priors, leading to divergent popularity trajectories, precisely the path-dependence that Bayesian social learning predicts, but now embedded in an active sampling framework where listeners choose which songs to explore based on expected free energy. The active sampling formulation adds a crucial dimension: it explains why social influence does not produce complete convergence. When a song's social prior is very strong, sampling it provides high pragmatic value but low epistemic value (the outcome is already largely predicted). Listeners with high epistemic drive may therefore resist the cascade, seeking less popular but informationally richer alternatives, a mechanism that generates the heterogeneity in cascade vulnerability that purely observational models struggle to explain.

The mere exposure effect, recast as a prediction effect (Huron, 2006), describes the progressive updating of the listener's generative model through repeated encounters. Each exposure enables the model to encode additional statistical regularities, increasing processing fluency. The inverted-U relationship between exposure and preference maps onto a specific model updating dynamic: early exposures provide high epistemic value (the model is substantially revised with each encounter), producing increasing liking as the song becomes predictable enough to track but still generates informative prediction errors. Overexposure, however, leads to convergence where the song generates essentially zero prediction error (the model has fully learned the stimulus), and only diminishing pragmatic value remains. The precise mechanism through which this operates is formalized in Section 5.1 in terms of Dirichlet concentration parameter accumulation on the likelihood mapping.

This mechanism has an important implication for cultural markets that standard models miss: the optimal number of exposures is not a fixed quantity but depends on the complexity of the song relative to the listener's current model. A harmonically complex song provides epistemic value over more exposures than a simple one; a listener with a more sophisticated generative model (higher cultural capital) can extract epistemic value from a given song for longer before reaching satiation. This means that the same recommendation algorithm, serving the same number of repeated plays, will produce different preference trajectories for listeners with different model complexities, a prediction with direct consequences for the personalization-versus-diversity debate in platform design.

Streaming algorithms function as niche-constructing agents that reshape the musical environment. The algorithm performs active inference on the listener's behalf, selecting observations predicted to minimize the listener's free energy, a form of epistemic exploitation that samples from well-calibrated regions of the musical space. The result, as Anderson et al. (2020) documented for Spotify, is systematic diversity reduction: the algorithm learns what the listener already likes and provides more of it, progressively narrowing the range of musical encounters.

However, the dynamics are platform-specific. TikTok's virality-optimizing algorithm can introduce genuinely novel content (as evidenced by the breakthrough of Brazilian regional genres and K-pop acts through the platform), whereas Spotify's personalization tends toward comfort zone maintenance. This heterogeneity in algorithmic architectures creates differentiated niche construction effects across platforms, a nuance absent from the monolithic 'filter bubble' narrative. We use the term 'niche construction' in its active inference sense (Constant et al., 2019), referring to the modification of the agent's sensory environment through its own actions, a formal extension of the biological concept to cognitive and cultural domains. What makes algorithmic niche construction distinctive is its *scale and speed*: while an individual listener's own choices modify her environment slowly (she can only listen to so much music per day), algorithmic mediation accelerates and amplifies the feedback loop, making the transition dynamics in the formal model non-stationary on a timescale much shorter than would occur through unmediated consumption.

Bourdieu's habitus corresponds to structural priors initializing the generative model during critical developmental periods. Higher cultural capital yields a more diverse and complex set of structural priors, enabling epistemic value across a wider range of musical encounters. We specify this precisely: cultural capital broadens the prior preference function in two ways. It shifts the peak toward higher complexity (requiring more substantial prediction error for epistemic reward) and it widens the function (making a broader range of musical encounters rewarding). These are empirically distinguishable predictions: the first implies that musically educated listeners should prefer more complex music; the second implies that they should enjoy a wider *variety* of music. The cultural omnivore phenomenon is more consistent with the second prediction, suggesting that the primary effect of cultural capital is broadening the range of rewarding encounters rather than simply raising the complexity threshold.

A song achieves popularity when it satisfies three conditions simultaneously: *(i) optimal prediction error* for the target population's generative models; *(ii) learnable structure* enabling efficient model updating within a few listenings; and *(iii) social amplification compatibility:* features that make the song easy to share, discuss, and perform socially. These conditions interact in convergent feedback loops: social amplification increases exposure, which shifts optimal prediction error ranges through model updating; algorithmic curation selects for songs satisfying (i) and (ii) for the largest number of listeners, further amplifying the successful and suppressing the marginal. The convergent dynamics through which these individual-level mechanisms produce population-level outcomes, including the conditions under which they lead to ecosystem impoverishment, are formalized in the following section.

## 5. Formal Model and Computational Simulation

### 5.1 POMDP Specification

We formalize the integrative account of Section 4 as a partially observable Markov decision process (POMDP), a standard framework for sequential decision-making under uncertainty, widely used in computational economics, operations research, and artificial intelligence (Smith, Friston & Whyte, 2022; Da Costa, Parr, Sajid, Hesp, Smith & Friston, 2020). The POMDP is well suited to cultural consumption because it naturally accommodates the features that make cultural markets distinctive: agents must choose among goods whose true characteristics they cannot directly observe (partial observability), each choice updates both the agent's knowledge and the environment from which future choices are drawn (learning and niche construction), and the agent trades off exploiting familiar goods against exploring unfamiliar ones (the exploration–exploitation balance that is central to the cultural capital literature).

**Hidden states** ($s$). A listener agent $i$ at time $t$ faces hidden states partitioned into $s^{mus} \in \mathbb{R}^d$ (a song's position in a $d$-dimensional musical feature space capturing harmonic content, timbre, rhythm, and form) and $s^{soc} \in \{$popular, non-popular$\}$ (the song's social standing). The agent cannot directly observe $s^{mus}$ before listening; she observes only the auditory signal that listening produces.

**Observations and the likelihood mapping** ($o$, **A**). Upon selecting song $j$, the agent receives an auditory observation $o_j$. The likelihood mapping **A** specifies the probability of observations given hidden states. We model **A** as a categorical distribution with learnable parameters: for each song $j$ and observation category $k$, the likelihood is

$$A_{kj} = P(o = k \mid s = j) = a_{kj} / \Sigma_l \, a_{lj} \qquad (1)$$

where $\mathbf{a} = \{a_{kj}\}$ are *Dirichlet concentration parameters*. This is the standard conjugate-exponential parameterization for categorical likelihoods under active inference (Smith, Friston & Whyte, 2022, Section 4): the columns of **A** are drawn from Dirichlet distributions Dir($\mathbf{a}_j$), and the concentration parameters serve as sufficient statistics encoding the agent's accumulated experience with each song.

The key property is the *learning rule*. When agent $i$ listens to song $j$ at time $t$ and receives observation $o = k$, the concentration parameters update as:

$$a_{kj}(t + 1) = a_{kj}(t) + 1 \qquad (2)$$

All other entries of **a** remain unchanged. This is the standard Bayesian update for a Dirichlet–categorical model: each exposure increments the relevant count by one. The consequence is that the column $\mathbf{a}_j$ grows with each exposure, and the corresponding likelihood $A_{kj}$ becomes increasingly peaked: the agent's predictions about what song $j$ will sound like become more precise.

**Surprisal.** From the likelihood mapping we derive the key quantity linking **A** to the preference function. The *surprisal* (or information content) of observation $o_j$ under the agent's current model is:

$$I_i(o_j, t) = -\log P_i(o_j, t) = -\log \Sigma_s \, A\{o_j, s\}(t) \cdot Q_i(s, t) \qquad (3)$$

where $Q_i(s, t)$ is the agent's current posterior over hidden states. As concentration parameters $a_{kj}$ accumulate through Eq. (2), the likelihood A becomes more peaked, the marginal probability $P(o_j)$ increases (the observation becomes more predictable), and surprisal $I$ decreases. This monotonic decline of surprisal with exposure is the formal mechanism underlying the mere exposure effect: listening to a song makes it more predictable.

**Prior preferences** ($C$). The preference function $C$ encodes which observations the listener finds rewarding, not in terms of specific songs, but in terms of the *degree of surprise* they generate. To capture the inverted-U relationship between novelty and pleasure documented in Section 2.2, we define $C$ as a function of surprisal:

$$C(o) = -(I(o) - \mu_c)^2 / (2\sigma_c^2) \qquad (4)$$

where $I(o)$ is the surprisal from Eq. (3), $\mu_c$ is the listener's *preferred surprisal level* (the sweet spot), and $\sigma_c$ is the *tolerance bandwidth*. This Gaussian specification generates maximal pragmatic value when a song's surprisal matches the listener's preferred level, declining symmetrically for songs that are too predictable ($I(o) < \mu_c$) or too surprising ($I(o) > \mu_c$).

Because $I(o)$ depends on the Dirichlet parameters **a** via Eqs. (1)–(3), and **a** updates with each exposure via Eq. (2), the preference landscape $C$ is implicitly *state-dependent*: it shifts as the agent learns. This is a non-standard feature of the POMDP (where $C$ is typically a fixed vector), but it is precisely what implements the endogeneity of preferences that distinguishes active inference from rational addiction. As the listener's model sharpens through repeated exposure, the same song generates less surprisal, moves through the $\mu_c$ sweet spot, and eventually falls below it. Preferences change because the evaluative apparatus changes, not because a consumption capital stock grows as the utility function remains fixed.

The interaction of Eqs. (2)–(4) generates three qualitatively distinct listening trajectories, depending on where a song's initial surprisal falls relative to the listener's $\mu_c$:

(i) *Inverted-U* (initial $I > \mu_c$): The song starts as too surprising; Dirichlet accumulation via Eq. (2) gradually reduces $I$ through the sweet spot (rising limb of $C$) and eventually below it (descending limb). The peak occurs later for more complex songs.

(ii) *Monotonic decline* (initial $I$ slightly below $\mu_c$): The song starts as somewhat too predictable; each exposure drives $I$ further below the sweet spot without ever passing through it.

(iii) *Never rewarding* (initial $I \ll \mu_c$): The large quadratic penalty in Eq. (4) cannot be offset by epistemic value, even on first encounter.

These three trajectory types, jointly predicted by the interaction of initial surprisal, Dirichlet learning, and the $C$ function, constitute a novel empirical signature of the model.

Cultural capital shapes $C$ in two ways: it increases $\mu_c$ (shifting the sweet spot toward higher complexity) and it increases $\sigma_c$ (broadening the bandwidth). The cultural omnivore phenomenon is more consistent with the second prediction, suggesting that the primary effect of cultural capital is broadening the range of rewarding encounters rather than simply raising the complexity threshold. We model cultural capital through a scalar $\sigma_c$; actual cultural capital is domain-specific, and future extensions should explore vector-valued $C$ functions representing genre-specific competences.

**Expected free energy and policy selection.** The agent selects the policy $\pi$ that minimizes *expected free energy* $G(\pi)$. For a policy that involves listening to song $j$, $G$ decomposes as (Da Costa et al., 2020):

$$G(\pi_j) = E_{Q(o|j)} [\log Q(s) - \log P(o, s)] \qquad (5a)$$

which can be rearranged into two interpretable terms:

$$G(\pi_j) = -D_{KL}[Q(s \mid o, j) \parallel Q(s)] - E_{Q(o|j)}[\log \tilde{P}(o)] \qquad (5b)$$

$$= -(\text{epistemic value}) - (\text{pragmatic value})$$

The first term, *epistemic value*, is the expected Kullback–Leibler divergence between the posterior $Q(s \mid o, j)$ after hearing song $j$ and the current prior $Q(s)$: the expected information gain from sampling. Songs the agent has heard many times have well-calibrated likelihoods (high concentration parameters in Eq. 1), so the posterior will barely differ from the prior: epistemic value is low. Novel songs, where $\mathbf{a}_j$ is small and the likelihood is diffuse, offer high expected posterior revision: epistemic value is high.

The second term, *pragmatic value*, is the expected log probability of preferred observations, where $\tilde{P}(o) \propto \exp(C(o))$ encodes the preference function from Eq. (4). Songs near the $\mu_c$ sweet spot have high pragmatic value; songs far from it have low pragmatic value regardless of their novelty.

In musical terms: epistemic value captures the reward of *learning something new*; pragmatic value captures the hedonic reward of *hearing preferred music*. Their relative weighting governs whether the listener explores or exploits, and this weighting changes dynamically as $\mathbf{a}$ accumulates: the exploration–exploitation balance is endogenous rather than externally imposed. Algorithmic curation systematically biases toward exploitation by constraining which songs are available for sampling (see **B** below), limiting the scope for epistemic-value-driven exploration.

A formal consequence of the state-dependence of $C$ deserves acknowledgment. When $C$ is fixed, the two terms in Eq. (5b) are cleanly separable. When $C$ depends on the posterior through $I(o)$ and hence $\mathbf{a}$ (as it does here via Eqs. 1–4), updating $Q(s)$ simultaneously shifts the pragmatic value landscape. The decomposition holds exactly within a single decision step (no parameters have yet updated) and is approximately valid across the short planning horizons relevant to song-by-song selection (in our simulation below, for instance, agents select 5 songs per session, and the per-exposure Dirichlet update is small relative to accumulated concentration).

**Transition dynamics and algorithmic curation (B).** In the standard POMDP, the transition matrix $\mathbf{B} = P(s_{t+1} \mid s_t, \pi)$ governs how hidden states evolve. For cultural markets, the critical feature is that the agent's *feasible choice set* at time $t+1$ depends on aggregate behavior at time $t$: algorithmic curation determines which songs are visible. Rather than specifying **B** as a full state-transition matrix (which would be intractable for the continuous feature space), we operationalize the policy-dependent component of **B** as an *availability pool*: a structural constraint on the set of songs from which the agent can sample:

$$L(\alpha) = \text{round}(\text{pool}_{size} \cdot (1 - \text{pool}_{shrink} \cdot \alpha)) \qquad (6)$$

where $\alpha \in [0, 1]$ is the *effective curation parameter*. Of the $L(\alpha)$ visible songs, approximately $\alpha \cdot L(\alpha)$ are drawn from a shared popularity-ranked list (increasing exposure overlap across listeners as $\alpha$ rises), and the remainder are similarity-sampled from the listener's neighborhood in feature space (with a small discovery mass ensuring non-zero exposure for all songs).

This implements algorithmic influence through a constraint on the *action space* (which songs can be sampled) rather than through direct manipulation of preferences or utilities, a distinction that matters for policy analysis, since it means that curation operates by limiting epistemic access rather than by distorting evaluation.

**Initial state beliefs (D).** The prior $D = P(s_0)$ encodes beliefs about the musical environment before observation. Social signals shape $D$ by increasing precision on specific songs: a song with high social validation has a sharply peaked prior, biasing policy selection toward sampling it. In streaming markets, aggregate statistics (charts, stream counts) serve as the relevant social signals.

**Per-step dynamics.** We can now state the complete temporal structure. At each time step $t$, for each agent $i$:

Step 1. The platform constructs the visibility pool $P_{i,t}$ via Eq. (6), using the current cumulative play counts.

Step 2. The agent evaluates $G(\pi_j)$ from Eq. (5b) for each song $j \in P_{i,t}$, using the current concentration parameters $a_i(t)$, preference parameters ($\mu_{c,i}, \sigma_{c,i}$), and social influence from $D$ (cumulative play counts).

Step 3. The agent selects songs by sampling from a softmax distribution over $-G(\pi_j)$ (i.e., choosing songs with the lowest expected free energy, equivalently the highest combined epistemic and pragmatic value).

Step 4. For each selected song $j$, Dirichlet parameters update via Eq. (2): $a_{kj}(t+1) = a_{kj}(t) + 1$. Cumulative play counts $n_j$ are incremented.

Step 5 (supply side). Songwriters observe the updated play counts and, with probability equal to the conformity parameter, shift their production centers toward the popularity-weighted centroid of the feature space. This is the songwriter drift process detailed in Section 5.3.

The updated play counts feed back into Step 1 at the next time step (the popularity-ranked list changes), creating the policy-dependent transition channel: what the agent listened to at $t$ alters what is available at $t + 1$.

**Mapping the four mechanisms.** Each mechanism of Section 4 now maps to a specific POMDP element and equation: *Mechanism 1* (social priors) → $D$ (precision-weighted initial beliefs); *Mechanism 2* (exposure-dependent learning) → $a$ (Dirichlet update, Eq. 2); *Mechanism 3* (algorithmic niche construction) → availability pool (Eq. 6, the policy-dependent constraint on the action space); *Mechanism 4* (cultural capital) → $C$ (parameters $\mu_c$ and $\sigma_c$ in Eq. 4).

## 5.2 From Individual Models to Population Dynamics: Multi-Agent Extension

The most interesting predictions concern population-level dynamics, yet the POMDP formalism describes individual agents. Bridging this gap requires engaging with the multi-agent active inference literature, following Vasil, Badcock, Constant, Friston, and Ramstead (2020) for cooperative communication and Veissière, Constant, Ramstead, Friston, and Kirmayer (2020) for cultural learning.

The key concept is that of a shared generative model or 'regime of attention' (Ramstead, Veissière & Kirmayer, 2016). Constant et al. (2019) formalized this as a process in which agents have an *adaptive prior for alignment*: a prior preference for sensory evidence indicating that their mental states are similar to those of conspecifics. In the musical context, this manifests as the well-documented tendency for social groups to converge on shared preferences. In streaming markets, aggregate statistics (charts, stream counts) serve as the large-scale analog of direct behavioral observation, with **D** encoding socially transmitted priors about which songs are worth sampling.

Population-level convergence then emerges through the interaction of three coupled processes. First, *model alignment*: agents' adaptive prior drives individual generative models toward similarity, narrowing the population distribution of ($\mu_c$, $\sigma_c$) vectors. Second, *supply-side optimization*: songwriters observe the population's narrowing sweet spot through play counts and produce songs increasingly concentrated within it, creating a feedback loop in which narrowed supply trains listeners on restricted samples, further narrowing their $C$ distributions. Third, *algorithmic amplification*: curation algorithms detect and reinforce the emergent consensus through the popularity-ranked component of the pool (Eq. 6), preferentially surfacing already-popular content. The coupled three-process system constitutes a convergent feedback loop tending toward a low-entropy, low-epistemic-value equilibrium: the condition of musical impoverishment.

**Predictions.** The formalized model generates four testable predictions. First, consumption entropy $H_{\text{cons}}$ should be lower under stronger algorithmic curation (higher α in Eq. 6). Second, populations with more homogeneous $C$ distributions (lower $\sigma_c$ in Eq. 4) should exhibit steeper winner-take-all dynamics. Third, exogenous injection of novel content should produce a temporary resurgence of epistemic value (the first term of Eq. 5b), as novel songs with small **a**$_j$ offer high expected information gain. Fourth, listeners with higher $\sigma_c$ (broader preference bandwidth, i.e., higher cultural capital) should maintain higher consumption diversity over time, resisting algorithmic narrowing.

## 5.3 Case Selection and Parameter Calibration

The formal model generates predictions about how institutional structures shape ecosystem dynamics. To test whether these predictions are consistent with observable cross-national variation, we select four national music ecosystems that maximize institutional contrast along the dimensions identified in Section 5.2: the number of independent generative sources ($K$), the incentive structure governing supply-side conformity, the degree of algorithmically mediated exposure overlap (α), and the breadth of listener preference distributions ($\sigma_c$). We draw on the Skoove/DataPulse (2025) analysis of Spotify streaming data across 73 countries for comparable cross-national measures. For each case, we describe the institutional structure and then state the parameter values it motivates.

### Italy (Festival di Sanremo): convergent feedback loop

Italy's Festival di Sanremo provides a near-ideal case of the convergent dynamics predicted by the model. The festival functions as a closed feedback loop operating on an annual cycle: songs are composed and selected under strong expectations about audience acceptance; audience responses (televoting, streaming surges, social media) update songwriter and artistic director models of what 'works'; and the following year's entries are produced within an increasingly narrow window of the population's collective generative model. The structural incentives are precise: songwriters who deviate too far from the expected aesthetic risk elimination in early rounds, while those who track the consensus are rewarded with advancement, post-festival commercial success, and future commissions. The gatekeeper (the artistic director) faces analogous incentives: selections that produce audience rejection threaten ratings and institutional reputation.

The concentration of songwriting in a small professional circle is not a recent phenomenon but a structural feature of the Sanremo ecosystem that has been denounced by critics and industry observers for years. The 2025 edition, however, made the pattern quantitatively visible. An investigative article in *Il Sole 24 Ore* (21 January 2025) documented that just eleven authors had signed approximately two-thirds of the thirty songs in the competition, with the most prolific individual credited on seven entries and several others appearing on four or five each, thus creating a situation in which entries that audiences perceived as independent creative expressions in fact shared underlying generative sources. The consumer association Codacons subsequently filed a complaint with the Antitrust authority, requesting an investigation into what it termed a 'discography caste.' This is a structural feature, not merely an ethical failing: when the population of active songwriters is small and interconnected, independent generative diversity cannot be maintained regardless of the number of entries in competition. The gender homogeneity of this circle (overwhelmingly male) further constrains the aesthetic range entering the system. Research on creative industries consistently shows that demographic homogeneity among producers narrows the range of outputs (Kim, 2025; Lee & Kim, 2020), and this effect compounds over iterations as each year's narrowed output becomes the training data for the next generation of songwriter models.

*Parameter configuration.* **K = 3** (independent sources): the songwriter cartel evidence suggests that the effective number of independent creative centers is very small, even if the nominal number of entries is larger. **Conformity = 0.90**: the elimination format and post-festival commercial rewards create strong incentives to track the consensus, producing high supply-side drift toward the popularity-weighted centroid. **α = 0.95**: the festival is a nationally synchronized broadcast event that concentrates attention on a small shortlist of tracks; Italy is also a strongly domestic-leaning market in streaming (83% local content; Skoove & DataPulse Research, 2025), and the festival's media ecosystem further increases common exposure. **$\bar{\mu}_c = 1.0$, $\bar{\sigma}_c = 0.7$**: a narrow preference bandwidth reflects the restricted aesthetic range that the closed feedback loop produces over time: listeners whose models have been trained predominantly on Sanremo-adjacent output will have correspondingly narrow tolerance for surprisal.

### Brazil: structural polyphony

Brazil represents the polar opposite of Sanremo's convergent dynamics. The country maintains extraordinary generative diversity through the coexistence of samba, bossa nova, MPB (Música Popular Brasileira), forró, sertanejo, axé, funk carioca, pisadinha, and numerous other genres, each sustained by distinct regional production infrastructure rooted in the deep historical interaction of Afro-Brazilian, Portuguese, and indigenous musical substrates. Local streaming content is distributed across dozens of genres, a consumption pattern consistent with multiple independent generative sources maintaining high supply-side entropy.

The regional rootedness of Brazilian genres is critical for the model: because different genres are produced by different communities in different regions with different musical traditions, the independence of generative sources is *structurally maintained* rather than requiring institutional enforcement. A successful sertanejo songwriter faces no incentive to imitate funk carioca, because the audiences, reward structures, and production communities are distinct. TikTok and YouTube have further diversified access channels, enabling regional genres like pisadinha (originating in Bahia and Sergipe) to reach national audiences by bypassing the traditional gatekeeping infrastructure of São Paulo-based labels and broadcasters.

*Parameter configuration.* **K = 15**: a deliberately high number reflecting the many genuinely independent regional genre traditions, each with its own production infrastructure.

**Conformity = 0.02**: genre-specific audience expectations and reward structures minimize cross-genre imitation; the conformity parameter is very low because there is no single consensus to track. **α = 0.30**: despite a strong domestic footprint (Spotify reports that Brazilian rights-holders account for 84% of Daily Top 50 tracks; Spotify, 2025, whereas the Skoove & DataPulse data is not available), the coexistence of multiple large regional ecosystems limits the homogenizing pressure of platform curation. **μ̄$_c$ = 1.2, σ̄$_c$ = 1.2**: a broad preference bandwidth reflects the wide aesthetic tolerance fostered by immersion in a polyphonic musical environment, as listeners accustomed to genre diversity develop generative models that reward a broader range of surprisal levels.

### South Korea: competitive diversification under oligopoly

K-pop presents a structurally distinctive case that challenges the assumption that market concentration necessarily produces cultural homogeneity. The industry is dominated by a small number of entertainment companies (Hybe, SM Entertainment, JYP Entertainment, YG Entertainment) yet produces remarkable sonic diversity. The mechanism is competitive differentiation: because each company maintains multiple active groups whose commercial viability depends on distinctive sonic and visual identities, imitation between labels would be commercially suicidal, as it would cannibalize existing market segments rather than opening new ones. The multi-label system thus maintains independent generative models within the industry, a form of organized diversity. The trainee system draws from a large, demographically diverse pool, sustaining supply-side variety.

Kim (2025), analyzing category shifts in Korean popular music, found that high-status agencies are more likely to make radical stylistic shifts, suggesting that market power enables rather than constrains experimentation. International songwriter integration with producers from Sweden, the UK, and the US contributing to K-pop productions further diversifies the generative sources. However, synchronized release patterns, intense chart competition, and the concentration of attention on comeback cycles generate substantial exposure overlap across the listening population.

*Parameter configuration.* **K = 8**: more than Sanremo (genuine competitive independence between labels) but fewer than Brazil (the industry is concentrated in a handful of corporate entities). **Conformity = 0.10**: competitive differentiation incentivizes distinctiveness, keeping conformity low. **α = 0.65**: the high domestic chart presence (77% local acts in the Spotify Top 200; Skoove & DataPulse Research, 2025) and synchronized release patterns increase common exposure, but the diversity of label identities prevents the full concentration observed in Sanremo. **μ̄$_c$ = 1.1, σ̄$_c$ = 1.1**: intermediate preference bandwidth reflecting a culture of active fandom that rewards sonic innovation within idol groups, maintaining strong genre boundaries.

### United Kingdom: supply-side diversity with demand-side vulnerability

The UK case reveals a distinction that the model makes precise: the critical difference between supply-side diversity (how many independent generative sources produce music) and demand-side resilience (whether locally produced music captures a substantial share of consumption). The UK's production ecosystem exhibits exemplary diversity: grime, drum and bass, indie rock, electronic music, folk revival, and numerous subgenres, supported by independent labels, BBC radio, and a historically robust music press. Hesmondhalgh's (2019) comprehensive analysis documents how this infrastructure has generated disproportionate global influence relative to market size. By any supply-side measure, the UK is among the most generatively diverse ecosystems in the world.

Yet British music accounts for only 29% of UK streaming, compared with 55% American content (Skoove/DataPulse, 2025). The paradox of a country with world-class productive diversity that nonetheless consumes predominantly imported music resolves within the model. American music benefits from the largest social priors **D** in the English-speaking world (global cultural presence creating strong precision-weighted expectations that bias sampling toward American content) and from the most data-rich algorithmic optimization of **B** (the largest English-language market provides the most training data for recommendation algorithms, producing tighter personalization loops). The UK's traditional demand-side mechanism with the music press functioning as an 'epistemic value amplifier' directing attention toward novel music has weakened in the streaming era, replaced by algorithmically generated playlists that tend toward exploitation. The UK Music (2023) annual report documents a declining market share for domestic content alongside stable production output, providing empirical evidence for this supply–demand decoupling.

*Parameter configuration.* **K = 12**: reflecting the vibrant independent sector and geographic diversity of production centers (London, Manchester, Glasgow, Bristol). **Conformity = 0.30**: genre diversity and the independent ethos of the UK scene keep supply-side conformity relatively low, but not as low as in Brazil or South Korea where specific geographical or competitive differentiation incentives are at work. **α = 0.96**: the high reliance on a globally dominant repertoire (US artists occupy over half the UK Spotify Top 200; Skoove & DataPulse Research, 2025) naturally increases cross-user overlap on a shared set of global hits, producing Sanremo-like effective curation despite very different supply conditions. **$\bar{\mu}_c$ = 1.0, $\bar{\sigma}_c$ = 1.0**: moderate preference bandwidth as the UK has strong musical education infrastructure but the dominance of mainstream streaming playlists limits the population-level breadth that this education might otherwise produce.

## Summary

Table 1 collects the parameter values motivated above. The shared parameters (N = 200, M = 80, T = 60, γ = 8.0, β = 0.3, λ = 0.5, ω = 0.45, pool$_{size}$ = 18, pool$_{shrink}$ = 0.45) are held constant across scenarios to isolate the effects of institutional variation.

| Parameter | Sanremo | Brazil | K-pop | UK |
| --- | --- | --- | --- | --- |
| *K (independent sources)* | 3 | 15 | 8 | 12 |
| *Conformity* | 0.90 | 0.02 | 0.10 | 0.30 |
| *α (effective curation)* | 0.95 | 0.30 | 0.65 | 0.96 |
| *$\bar{\mu}_c$ (preferred surprisal)* | 1.0 | 1.2 | 1.1 | 1.0 |
| *$\bar{\sigma}_c$ (tolerance bandwidth)* | 0.7 | 1.2 | 1.1 | 1.0 |

*Table 1.* Simulation parameters for four institutional structures.

**Empirical anchoring of α.** Although α is not directly observed, it is interpretable through measurable proxies: (i) exposure overlap across users (e.g., Jaccard similarity of top-*N* recommended tracks); (ii) the share of listening mediated by salient playlists and editorial placements; and (iii) the degree to which nationally synchronized events concentrate attention. Evidence that User Interface (UI) salience materially shifts consumption supports treating salience-induced overlap as a key structural mechanism (Rackowitz & Haampland, 2025; Pachali et al., 2025). The values in Table 1 are intended to be consistent with the qualitative institutional differences described above, not as precise country-level estimates.

## 5.4 Computational Simulation

To verify that the predicted dynamics emerge from the structural properties specified above, we implement an agent-based simulation. This section describes the simulation as a *reduced-form approximation* of the formal POMDP, making explicit where and how it simplifies the full specification of Section 5.1.

**The central simplification.** The full POMDP requires maintaining Dirichlet parameters **a** for each agent–song pair, computing marginal likelihoods (Eq. 3), and evaluating expected free energy (Eq. 5) over the full state space, a computationally demanding task for $N = 200$ agents and $M = 80$ songs over $T = 60$ steps. The simulation therefore replaces the Dirichlet-based surprisal (Eqs. 1–3) with a distance-based proxy and replaces expected free energy minimization (Eq. 5) with a composite utility that captures the same economic tradeoffs in reduced form.

**Setup.** $N = 200$ listener agents choose among $M = 80$ songs in a two-dimensional feature space $[0, 4]^2$. $K$ source centers (songwriters/labels) are drawn from Uniform$(0.5, 3.5)^2$; songs are assigned to sources in round-robin and placed at their source's center plus Gaussian noise ($\sigma = 0.3$ per dimension, clipped to $[0, 4]^2$). Each agent $i$ has a model center $\mathbf{x}_i \sim$ Uniform$(0.5, 3.5)^2$ and preference parameters $\mu_{c,i} \sim$ Normal$(\bar{\mu}_c, 0.2)$, $\sigma_{c,i} \sim$ Normal$(\bar{\sigma}_c, 0.15)$, both clipped below at 0.3. All exposure counts $e_{ij}$ and play counts $n_j$ are initialized to zero.

**Composite utility.** At each time step $t$, each agent selects 5 songs by sampling without replacement from a softmax over:

$$U_i(j) = \gamma \cdot [C_i(j) + E_i(j)] + \omega \cdot \ln(1 + n_j) \quad (7)$$

where the four terms implement the POMDP elements as follows.

*Pragmatic value* (reduced-form Eq. 4). $C_i(j) = -(\tilde{I}_{ij} - \mu_{c,i})^2 / (2\sigma_{c,i}^2)$, where $\tilde{I}_{ij}$ is the *effective surprisal*, proxied by the Euclidean distance between the agent's model center and the song's position, attenuated by a familiarity function:

$$\tilde{I}_{ij} = \|\mathbf{x}_i - \mathbf{s}_j\| / (1 + \lambda \cdot e_{ij}) \quad (8)$$

with familiarity decay rate $\lambda = 0.5$ and $e_{ij}$ as the cumulative exposure count. This is the simulation's reduced-form approximation of the Dirichlet mechanism (Eqs. 1–3): as exposure $e_{ij}$ grows, effective surprisal $\tilde{I}$ declines: the song moves through the preference landscape $C$ as the formal model predicts, but via a computationally tractable proxy. The distance $\|\mathbf{x}_i - \mathbf{s}_j\|$ plays the role of initial surprisal (unfamiliar songs are 'far' in feature space), while the denominator $(1 + \lambda \cdot e_{ij})$ plays the role of Dirichlet accumulation (exposure reduces effective distance).

*Epistemic value* (reduced-form first term of Eq. 5b). $E_i(j) = 1 / (1 + \beta \cdot e_{ij})$, with $\beta = 0.3$. This decays with exposure, capturing the diminishing information gain from re-hearing a known song. The decay rate $\beta$ is slower than the familiarity attenuation $\lambda$ (0.3 vs 0.5), reflecting the fact that information gain diminishes more gradually than perceptual novelty.

*Social influence* (reduced-form **D**) is modeled as $\omega \cdot \ln(1 + n_j)$, where $n_j$ is the cumulative play count across all agents and $\omega = 0.45$. The logarithmic form produces diminishing marginal social evidence, consistent with the informational structure of cascades.

*Algorithmic curation* (reduced-form Eq. 6). Before evaluating Eq. (7), the platform constructs the visibility pool $P_{i,t}$ exactly as specified in Section 5.1 (Eq. 6). Agents choose only from songs in their pool. This ensures that curation operates through the constraint on the action space, not through utility manipulation.

The scaling parameter $\gamma = 8.0$ governs the relative weight of cognitive evaluation (pragmatic + epistemic) versus social influence. Each agent draws 5 songs from $\Pr(j) \propto \exp(U_{i(j)})$, then $e_{ij}$ and $n_j$ are updated.

**Songwriter drift** (implementing supply-side convergence, Step 5 above). After all agents have selected, each of the $K$ sources drifts toward the popularity-weighted centroid $\mathbf{c} = \Sigma_j (n_j / \Sigma_k n_k) \cdot \mathbf{s}_j$. With probability equal to the conformity parameter, a source shifts its center 12% of the distance toward $\mathbf{c}$; songs belonging to that source shift at half the rate (6%), maintaining within-source spread. All $K$ sources drift toward the same global centroid, which in the high-conformity Sanremo condition (conformity = 0.8) produces rapid spatial convergence.

**Relationship to the formal POMDP.** Two simplifications need an explicit comment here.

First, the Euclidean distance proxy (Eq. 8) conflates two quantities that the formal model distinguishes. In the POMDP, surprisal $I(o)$ (Eq. 3) measures how unexpected an observation is under the current generative model (an epistemic quantity), whereas $C(o)$ (Eq. 4) evaluates how far that surprisal falls from the preferred level (a pragmatic quantity). Using distance from the agent's model center as the proxy simultaneously captures unfamiliarity and feature mismatch, collapsing the epistemic–pragmatic distinction. This is the most consequential simplification; interpretations should be read with this conflation in mind.

Second, of the three convergence mechanisms identified in Section 5.2 (model alignment, supply-side optimization, and algorithmic amplification) the simulation implements only the latter two. Agents' preference parameters ($\mu_{c,i}$, $\sigma_{c,i}$) are fixed at initialization; the alignment mechanism is not simulated. The results are therefore conservative: incorporating preference alignment would strengthen the convergence dynamics.

Before turning to the four predictions, we present the overall ecosystem trajectories produced by the simulation under the four institutional structures of Table 1. Figure 1 tracks four metrics over the 60 time steps of the simulation: consumption entropy (panel A), which measures how evenly listening is distributed across the song catalogue; the Gini coefficient of attention (panel B), which captures the degree of winner-take-all concentration; mean epistemic value (panel C), which reflects the information gain that listeners obtain from their current listening; and supply-side dispersion (panel D), which measures the mean distance between production sources in the feature space, an index of how much the songwriter drift process has compressed the supply. The four scenarios diverge rapidly. Brazil maintains the highest consumption diversity and the lowest concentration throughout, whereas Sanremo and the UK converge toward high concentration from different starting conditions: Sanremo through supply-side collapse, the UK through demand-side channeling. K-pop settles at intermediate levels on all four metrics, consistent with its institutional structure of competitive differentiation under oligopoly. These trajectories provide the context for the four specific predictions that follow.

**Figure 1. Ecosystem Dynamics Under Four Institutional Structures**

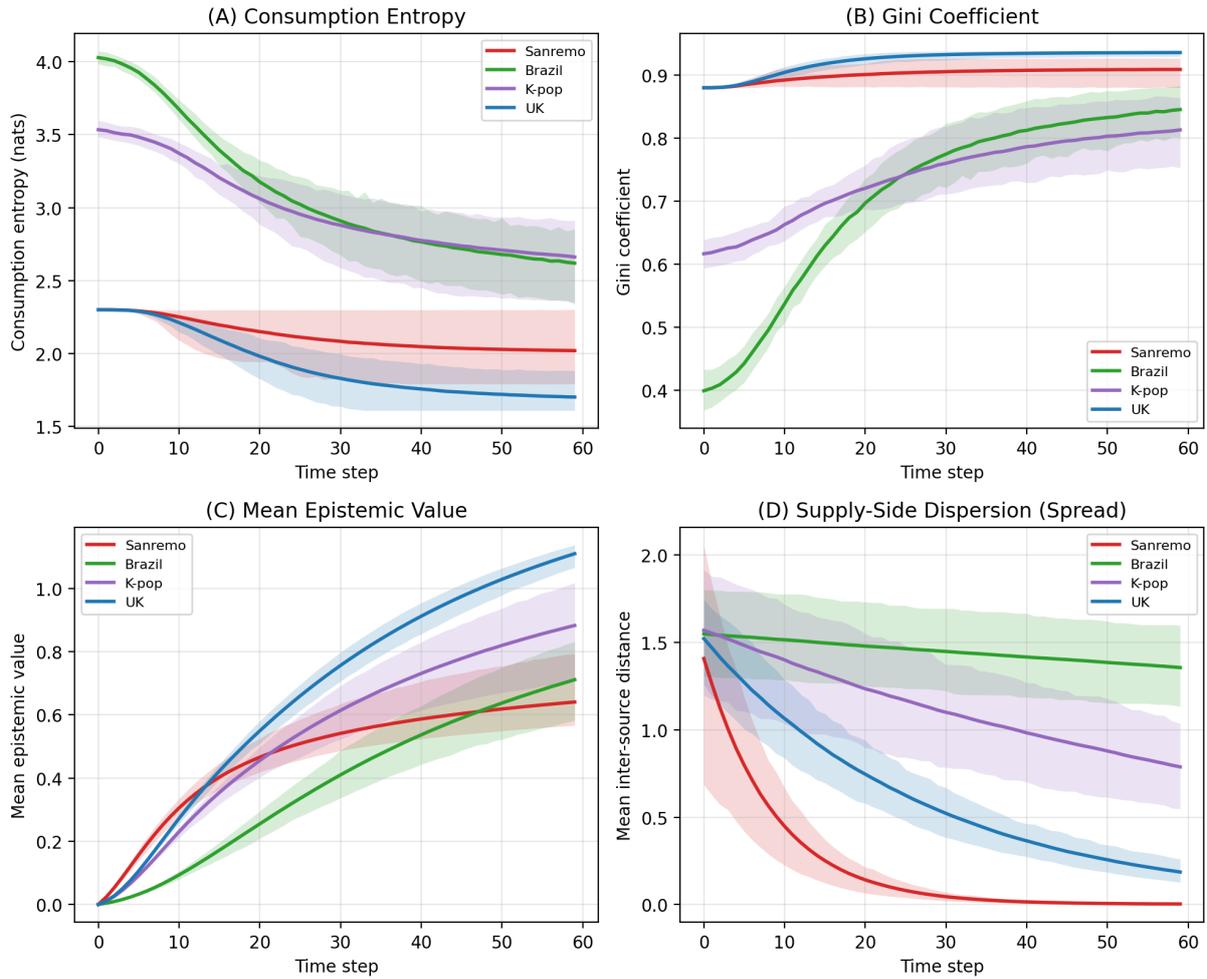

**Figure 1.** *Ecosystem dynamics under four institutional structures (R = 200, seed = 123, 5th–95th percentile bands). (A) Consumption entropy converges to the highest levels in Brazil and K-pop, remains lower in Sanremo, and is lowest in the UK. (B) Gini concentration is highest in the UK and Sanremo, and lower in Brazil and K-pop. (C) Mean epistemic value rises most strongly in the UK and K-pop, remains intermediate in Brazil, and is lowest in Sanremo. (D) Supply-side dispersion collapses toward zero in Sanremo, remains limited in the UK, and stays substantially higher in K-pop and especially Brazil. Parameters as in Table 1.*

**Results.** The simulation illustrates all four predictions.

*Prediction 1* (Figure 2A): Consumption entropy $H_{\text{cons}}$ decreases as α increases. The relationship is markedly nonlinear: entropy barely changes from α = 0.0 to 0.5 (≈4.31 to ≈4.21 nats, 2% decline), then drops sharply from α = 0.7 to 0.9 (≈4.04 to ≈3.12 nats, 23% decline). This threshold behavior suggests that moderate curation is relatively benign, but beyond a critical level the system tips into a self-reinforcing concentration regime.

*Prediction 2* (Figure 2B): Cross-national Gini coefficients reflect the joint effect of listener preference heterogeneity ($\sigma_c$) and institutional parameters (K, conformity, α). Even with moderate α, ecosystems with many sources and low conformity (Brazil) remain more egalitarian than systems with few sources and high conformity (Sanremo).

*Prediction 3* (Figure 2D): High cultural-capital listeners sustain higher individual consumption diversity, consistent with a buffer against homogenization under concentrated recommendation environments.

*Prediction 4* (Figure 2C and Supplementary Figure 3): High-curation, high-conformity systems collapse supply-side dispersion relative to pluralistic ecosystems.

**Figure 2. Four Robust Predictions Illustrated by Simulation**

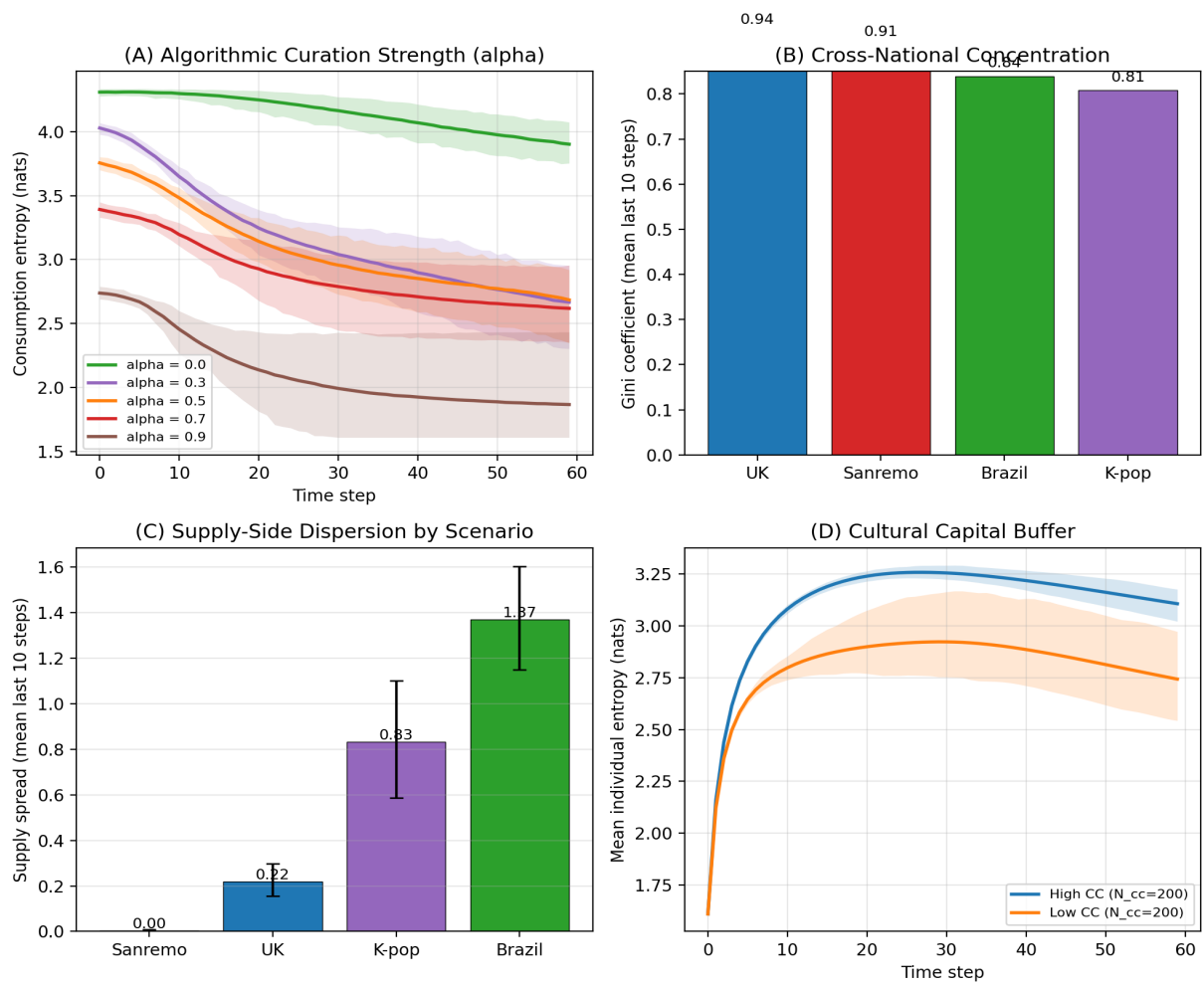

**Figure 2.** *Four robust predictions illustrated by the simulation (R = 200, 5th–95th percentile bands). (A) Consumption entropy decreases nonlinearly as algorithmic curation strength α rises, with a pronounced threshold beyond intermediate levels. (B) Cross-national concentration remains higher in the confirmatory contrasts that define Prediction 2: Sanremo − Brazil, UK − Brazil, and Sanremo − K-pop. (C) Supply-side dispersion remains near zero in Sanremo and much larger in Brazil, K-pop, and the UK. (D) High-cultural-capital listeners retain higher individual entropy than low-cultural-capital listeners under identical high-curation conditions.*

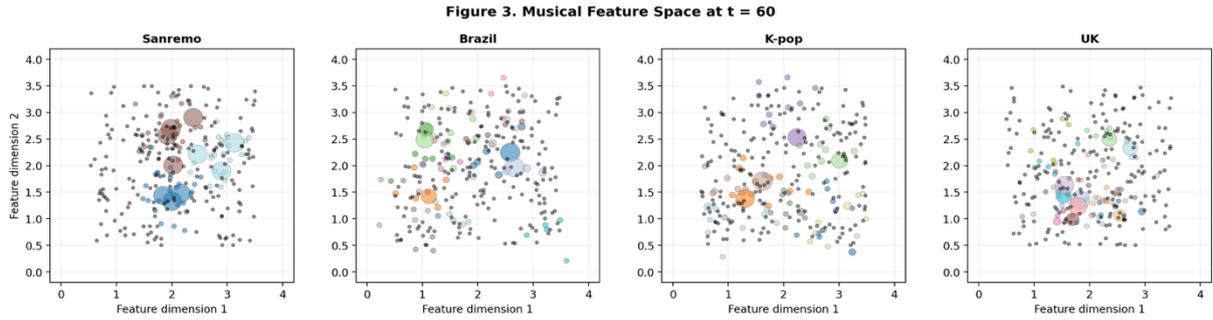

**Figure 3.** *Musical feature space at t = 60. Each panel plots songs in the two-dimensional feature space; circle size is proportional to cumulative play count; color indicates production source; grey points denote listener model centers. Sanremo shows tight convergence of the three source clusters; Brazil retains broad spatial dispersion across many independent sources; K-pop preserves intermediate clustering; and the UK remains more dispersed than Sanremo but markedly less spread than Brazil. These panels visualize the structural differences summarized numerically in Figure 4 of the replication package.*

**Supply-side convergence in feature space** (Figure 3). Figure 3 provides a vivid illustration of what we term *supply-side convergence* (or, more evocatively, 'feature space collapse'). At $t = 60$, Sanremo's songs have converged to a tight cluster in feature space, whereas Brazil's remain distributed across multiple distinct regions, the visual signature of structural polyphony versus songwriting cartel. We avoid the term 'precision collapse' because 'precision' carries a specific technical meaning in the active inference literature (the inverse variance of probabilistic beliefs), and what occurs here is the opposite: as songs converge spatially, the listener's generative model becomes *more* precise (higher confidence about what to expect), not less. The pathology is not that precision collapses but that the *object of precision* becomes impoverished: the model becomes very good at predicting a very narrow range of musical experiences. K-pop shows distinct clusters (competitive differentiation with $\text{supply}_{\text{spread}} = 0.70$), while the UK shows distributed songs but concentrated listener centers (demand-side vulnerability; $\text{supply}_{\text{spread}} = 0.66$). This visual impression is confirmed quantitatively by the $\text{supply}_{\text{spread}}$ metric (Figure 1D): Sanremo's mean inter-source distance collapses to effectively zero (0.006), indicating near-total spatial convergence of all three production sources, whereas Brazil maintains substantial separation (0.82).

**Robustness.** The four predictions are not artefacts of specific parameter choices. The qualitative ordering is preserved under $\gamma \in [3.0, 15.0]$; the monotonic relationship between $\alpha$ and entropy suppression holds across the full parameter range; and the Gini ordering between high-$\sigma_c$ and low-$\sigma_c$ populations is preserved under ±30% perturbation of each non-$\sigma_c$ parameter. However, the simulation is illustrative rather than estimative: its purpose is to show that the formal architecture generates the predicted dynamics, not to provide point estimates. Empirical calibration using streaming data and direct POMDP implementation with Dirichlet updating and expected free energy minimization remain priorities for future work.

Full numeric robustness evidence, including bootstrap confidence intervals from R = 200 replications, is reported in the Supplementary Information (Sections S2–S3).

## 6. Ecosystem Dynamics in Comparative Perspective

The simulation results, combined with the institutional analysis of Section 5.3, enable a richer interpretation of cross-national variation than either component provides alone. We discuss the four cases in the light of the model's predictions and identify four structural determinants of ecosystem vitality.

### 6.1 Sanremo: convergent impoverishment and curatorial disruption

The simulation confirms the institutional analysis: Sanremo's combination of few independent sources, high conformity, and strong curation produces near-total supply-side convergence (supply$_{spread}$ = 0.006) and the highest attention concentration (Gini = 0.62). But the simulation adds quantitative content that the institutional analysis alone cannot provide. The convergence is not merely a consequence of the cartel, it is *self-reinforcing*: even if the initial sources were more diverse, the high conformity parameter would drive them toward spatial convergence within approximately 20 time steps, and the high α would concentrate attention on the converging cluster. The songwriter cartel accelerates a process that the institutional incentive structure would produce regardless.

The consequences are empirically observable. ANSA reports the 2026 first night at 9.6 million viewers (58% share) versus 12.63 million (65.3%) in 2025; the 2026 final averaged 11.022 million (68.8%). This pattern is consistent with the model's prediction that low-epistemic, highly concentrated ecosystems truncate long-run engagement, though viewership is multiply determined.

The Amadeus years (2020–2024) provide a natural experiment that corresponds to Prediction 3 (novelty injection). Amadeus's selection of acts like Måneskin (punk rock energy alien to the Sanremo tradition) and Mahmood (Arabic-Italian vocal fusion drawing on cultural substrates entirely outside the festival's palette) generated substantial prediction error for the festival audience. However, such prediction error moved along the epistemic arc described by Van de Cruys et al. (2024): complete trajectories from curiosity through exploration to resolution. Viewership rose, both acts achieved international success, and cultural relevance increased. The reversal under Carlo Conti (2025–2026) to conservative curation represents a shift from exploration to exploitation in curatorial policy, and the declining engagement metrics are consistent with the simulation's finding that injection effects are temporary without sustained structural intervention.

The implication for public broadcasting is clear. Public broadcasters can absorb the individual-entry risk that commercial gatekeepers cannot: a publicly funded festival can include entries that initially challenge the audience, because its survival does not depend on maximizing each entry's audience response. The recommendation is not for all-in avant-garde provocation but for *structured novelty*: curatorial selection that generates curiosity, enables exploration, and rewards resolution.

We emphasize that the Sanremo analysis is *illustrative rather than evidential*. Systematic empirical testing should include a computational analysis of harmonic and timbral convergence using IDyOM (Pearce, 2005), formal HHI calculation of songwriting credits, and comparison of prediction error distributions across the Amadeus versus Conti periods.

### 6.2 Brazil: the healthy equilibrium

Brazil's simulation results, that is, the highest consumption entropy (≈4.12 nats), the lowest Gini (0.37), and sustained supply-side dispersion (0.82), confirm that the institutional features described in Section 5.3 jointly produce the polyphonic equilibrium that the model predicts. But again the simulation reveals something that the institutional analysis does not: the *mechanism* through which polyphony sustains itself. Because $K = 15$ sources are distributed across the feature space, no single popularity-weighted centroid can attract all of them; the drift process pulls different sources toward different local attractors rather than a single global one, maintaining spatial diversity even under nonzero conformity. This is a structural buffer that operates independently of any deliberate policy of diversity maintenance: it is a consequence of geographical and cultural distance among production centers.

The coexistence of high domestic consumption share (84%) and high generative diversity makes Brazil a 'polyphonic fortress' with high domestic engagement sustained by genuine variety, not by convergent concentration on a narrow national style. This configuration contrasts sharply with Italy's 'convergent fortress' (high domestic share masking low internal diversity), and the distinction is invisible to any model that measures only the domestic/international ratio without examining within-domestic entropy.

### 6.3 K-pop: organized diversity under oligopoly

K-pop's intermediate position in the simulation (Gini = 0.49, supply$_{spread}$ = 0.70) reflects its distinctive institutional structure: competitive pressure between labels maintains aesthetic differentiation, but synchronized release patterns and intense chart competition generate substantial exposure overlap. The simulation makes a specific prediction that the institutional analysis alone cannot: K-pop's ecosystem should be *more robust to increases in α* than Sanremo but *less robust* than Brazil, because its intermediate $K$ and low conformity provide some buffer against concentration while its moderate α already produces partial overlap. This prediction is testable through consumption entropy comparisons across Spotify markets with different playlist prominence.

Gender norms impose a structural limitation visible in the simulation's feature space (Figure 3): female groups face stronger constraints on sonic experimentation than male groups (Kim, 2025), reducing the effective dimensionality of one segment's feature space. The simulation's $K = 8$ treats all sources as equally free to explore; a more realistic specification would assign lower effective $\sigma_c$ to sources representing female groups, which would further concentrate one region of the feature space, leaving the male segment relatively dispersed.

### 6.4 The United Kingdom: the supply–demand paradox

The UK produces the most striking result in the simulation: supply$_{spread}$ = 0.66 (dispersed production) coexisting with Gini = 0.61 (nearly Sanremo-level attention concentration). The simulation nails down what the institutional analysis suggests: supply-side diversity is *necessary but not sufficient* for ecosystem vitality. The UK's high α = 0.82, driven not by a national festival but by the dominance of globally shared repertoire in algorithmic recommendations, channels consumption toward a narrow band of already-popular songs despite the availability of a highly diverse local production. The traditional demand-side corrective (the music press as 'epistemic value amplifier') has weakened in the streaming era; the UK Music (2023) annual report documents a declining domestic market share alongside stable production output, precisely the supply–demand decoupling that the model predicts when high α operates on a market with strong social priors **D** favoring imported content.

### 6.5 Four structural determinants of ecosystem vitality

The cross-national comparison identifies four structural determinants, each mapped to specific model elements:

**(i)** Independent generative sources → $K$ and supply-side entropy. The effective number of independent songwriting and production centers determines baseline variety. Brazil's regional traditions provide maximal independence; Sanremo's cartel provides minimal independence; K-pop's competitive labels occupy an intermediate position; the UK's vibrant independent sector provides high supply-side diversity ($K = 12$), though this diversity does not automatically translate into consumption diversity.

**(ii)** Institutional incentives for novelty → the $C$ function. The reward structure facing songwriters and curators determines whether the system incentivizes consensus-tracking (narrowing the population's $C$) or exploration (maintaining it). Sanremo's elimination format rewards consensus; K-pop's competitive differentiation rewards distinctiveness; Brazil's genre-specific audiences provide multiple distinct sweet spots; the UK's independent ethos historically rewards experimentation, but the weakening of the music press as an epistemic intermediary has reduced the institutional amplification of novelty.

**(iii)** Gatekeeping porosity → **B** and the availability pool (Eq. 6). High porosity, as in Brazil's TikTok/YouTube bypass of label gatekeepers, or K-pop's trainee pipeline, maintains supply-side entropy. Low porosity as in Sanremo's reliance on established songwriter networks accelerates convergence. The UK occupies a distinctive position: its production gatekeeping is highly porous (the independent sector is large and accessible), but its *consumption* gatekeeping is dominated by global streaming algorithms that channel attention toward imported content.

**(iv)** Demand-side resilience → **D** × **B** interaction. Markets with strong local institutions (Brazil's regional media, Korea's idol culture) resist algorithmic homogenization because local content generates sufficient engagement to shape algorithmic recommendations. Italy's high domestic share (83%) appears to indicate resilience, but masks low generative diversity within that share: a 'convergent fortress market.' The UK represents the opposite pathology: genuine supply-side diversity coexisting with demand-side vulnerability, as high α driven by globally dominant repertoire produces Sanremo-level attention concentration (*Gini* = 0.61) despite the rich local production. A direct comparison of within-domestic-share genre entropy between Italy and Brazil would provide the critical test of the convergent fortress characterization; we predict Italy's will prove substantially lower, but this remains an empirical claim awaiting verification.

The Bello and Garcia (2021) finding of increasing cross-national chart diversity is entirely consistent with this framework: global divergence reflects the growing between-country distinctiveness of robust ecosystems like Brazil and Korea and is compatible with within-country impoverishment in specific cases like Italy.

The fortress market concept requires case-by-case assessment. For instance, India's linguistic diversity with Hindi, Tamil, Telugu, Bengali, and Punjabi music industries operating quasi-independently likely makes it a 'polyphonic fortress' structurally analogous to Brazil. The framework generates this as a testable prediction rather than an assumption.

## 7. Cultural Policy Implications: Welfare Foundations

The preceding analysis shows that musical ecosystems can converge on impoverished equilibria through the interaction of social influence, institutional gatekeeping, and algorithmic curation. But showing that impoverishment *occurs* does not establish that it constitutes a *market failure* calling for policy intervention. The welfare case must be made explicitly, and it requires engaging with foundational questions about the normative status of culturally shaped preferences.

## 7.1 Three Welfare-Theoretic Foundations

We ground the case for intervention in three complementary frameworks. Each performs a distinct function in the argument. The *capability approach* provides the normative foundation: it specifies *what is at stake* and *why it matters*. The *adaptive preferences argument* provides the diagnostic function: it explains *why standard welfare analysis fails to detect the problem*. And the *externalities framework* provides the economic mechanism: it identifies *why markets underprovide the relevant goods*. No single framework is sufficient: capabilities without the adaptive preferences diagnosis could be dismissed as paternalism; the adaptive preferences argument without capability foundations lacks normative content; and the externalities argument without the other two reduces to a standard market failure that does not capture what is distinctive about cultural goods. Their convergence strengthens the normative argument beyond what any single framework could provide.

We begin with Sen's capability approach, which offers the normative foundation for our analysis. Sen (1985, 1999) argues that development should be evaluated not by commodity bundles consumed or by subjective utility reported, but by the substantive freedoms ('capabilities') that individuals enjoy to live the lives they have reason to value. What matters is not whether a person consumes a particular good but whether she has the effective opportunity to choose from a range of valuable functionings. The capacity for rich aesthetic experience, the ability to engage with diverse cultural stimuli, to experience epistemic arcs of curiosity, exploration, and resolution, to develop and refine one's generative model of musical structure across multiple traditions and idioms, constitutes a capability in Sen's sense.

The systematic erosion of this capability by feedback loops that narrow the available musical environment constitutes a *capability deprivation*. The deprivation is structural rather than coercive: no one is prevented from listening to diverse music in principle. But when the musical environment, shaped by algorithmic curation, supply-side convergence, and social influence dynamics, offers a progressively narrower range of aesthetic experiences, the effective capability set is reduced regardless of nominal freedom of choice. The analogy with nutritional capabilities is instructive: a person living in a food desert is not *prohibited* from eating nutritiously, but their effective capability to do so is severely constrained by the structure of their environment. Similarly, a listener in a musically impoverished ecosystem retains nominal access to global catalogues but faces an environment in which algorithmic curation, social priors, and supply-side convergence systematically direct attention toward a narrow band of familiar output.

A second key element is adaptive preferences, that provide the diagnostic function. Elster (1982), in his analysis of 'sour grapes,' established that preferences can adapt to constraints in ways that make the constraints invisible to revealed preference analysis. The fox who cannot reach the grapes declares them sour; the preference change is a response to the constraint, not an independent evaluation. This mechanism is not merely a philosophical possibility but a formal consequence of the active inference framework developed in Sections 3–5. Individuals in impoverished musical environments develop narrow preference functions $C$ (low $\sigma_c$, low $\mu_c$) because their generative models have been trained on a restricted sample. They come to *prefer* the impoverishment that constrains them, not through false consciousness or irrationality but through the entirely rational process of model optimization given available data. A listener whose generative model has been shaped by years of algorithmically curated listening has genuinely different preferences from one exposed to a richer environment, and the narrow preferences are genuinely experienced as satisfying.

This is the diagnostic insight that motivates the capability approach as the appropriate normative framework: because preferences are endogenous to the learning process (the central insight of Section 3.2.1), revealed preference cannot serve as an independent criterion for evaluating market outcomes. When the listener's generative model has been shaped by the very environment whose adequacy is in question, her expressed satisfaction with that environment cannot independently validate it. This is not paternalism in the standard sense: the claim is not to know better than the listener what she *should* prefer. We are pointing out that the preference formation process itself has welfare-relevant properties that standard welfare analysis ignores. The latter treats preferences as the measuring rod and market outcomes as the object of measurement; when the measuring rod is itself shaped by market outcomes, the circularity must be acknowledged. The capability approach escapes this circularity by evaluating outcomes against substantive freedoms rather than subjective satisfaction.

Crucially, the adaptive preferences diagnosis is not merely a theoretical possibility but receives empirical support from *reversibility evidence*. The Amadeus years at Sanremo (Section 6.2) provide a natural experiment: when audiences accustomed to formulaic output were exposed to genuinely novel acts (Måneskin, Mahmood), they did not reject the unfamiliar material; they responded with deepened engagement, rising viewership, and wide adoption. If the narrow preferences were 'deep' values rather than adaptive responses to a constrained environment, exposure to challenging material should have produced audience flight. The positive response suggests that the narrow preferences were indeed adaptive, shaped by the impoverished environment rather than reflecting stable underlying values, and that the capacity for richer aesthetic engagement was latent, requiring only an environmental change to manifest. This reversibility asymmetry (it is easier to broaden narrow preferences through exposure than to narrow broad ones) is a distinctive prediction of the active inference account: a generative model trained on a narrow sample can be updated by novel data, whereas a model trained on a rich sample does not 'forget' its complexity when exposed to simple material.

The economic mechanism is then provided by positive externalities. Musical diversity generates spillover benefits that are not captured by individual consumption decisions and therefore not reflected in market equilibria. These externalities operate through at least four channels: *innovation spillovers* (diverse musical ecosystems provide richer source material for creative recombination, both within music and across creative industries); *social cohesion effects* (shared but diverse cultural references enable richer social bonding and cross-group understanding than homogeneous ones); *identity formation* (the availability of diverse musical traditions enables individuals and communities to construct and maintain distinctive cultural identities, a function that is particularly salient for minority and immigrant communities); and *cross-domain creativity effects* (research in cognitive science suggests that exposure to diverse aesthetic structures enhances creative problem-solving and cognitive flexibility in non-musical domains). When individual consumption decisions, mediated by algorithmic curation optimized for individual engagement metrics, systematically reduce diversity, these public goods are undersupplied relative to the social optimum. The active inference framework gives this standard market failure argument particular force by specifying the mechanism through which individual-level optimization (minimizing each listener's free energy) produces a population-level outcome (ecosystem convergence) that no individual would choose if they could observe the system-level consequences of their aggregated choices.

## 7.2 Policy Implications

The welfare foundations yield four specific policy implications, each grounded in specific parameters of the formal model.

*In the first place, diversity of generative sources matters irrespectively of quality of individual outputs.* The formal model shows that ecosystem vitality depends primarily on the number of independent generative sources ($K$) and the entropy of their distribution $H_{supply}[P(s^{mus})]$, without a need to factor in any measure of individual song quality. Policies should focus on maintaining and expanding the infrastructure that supports independent production (regional studios, independent labels, training programs for diverse communities of songwriters, support for genre-specific production ecosystems) rather than on selecting, subsidizing, or promoting specific works meeting specific judgment criteria. The Brazilian case shows that structural diversity, not curatorial quality control, is the primary determinant of ecosystem health.

*Additionally, algorithmic curation requires diversity safeguards.* The simulation illustrates that stronger algorithmic curation monotonically suppresses consumption entropy (Prediction 1). Diversity safeguards, analogous in spirit to local content quotas in broadcasting but calibrated to the specific dynamics of streaming platforms, should ensure that recommendation algorithms maintain minimum levels of exploration alongside exploitation. Importantly, these safeguards must be platform-specific: TikTok's virality dynamics differ fundamentally from Spotify's personalization, and YouTube's session-length optimization produces yet different effects. A regulatory framework that treats all platforms identically would be poorly calibrated to actual dynamics. The emerging EU regulatory attention to algorithmic transparency in music streaming (European Parliament, 2024) provides a potential institutional basis for such differentiated approaches.

*Moreover, public institutions bear responsibility for epistemic richness.* The Amadeus/Conti contrast at Sanremo illustrates that curatorial choices by public institutions have measurable effects on ecosystem dynamics. Public broadcasters, festivals receiving public funding, and publicly supported cultural institutions should be understood as *epistemic value amplifiers*, institutions whose distinctive contribution is directing attention toward music that can generate complete epistemic arcs, maintaining supply entropy above the threshold that sustains positive epistemic value across the ecosystem. This reframes the standard justification for public cultural institutions from quality provision (which assumes stable exogenous preferences that the institution serves) to ecosystem maintenance (which recognizes that preferences are endogenously shaped by the environment that the institution helps to create).

*Finally, musical education as prior enrichment.* In the formal model, cultural capital operates by broadening the preference function $C$, increasing both the preferred complexity level ($\mu_c$) and the tolerance bandwidth ($\sigma_c$). Music education programs function as systematic prior enrichment, expanding the population's capacity for diverse aesthetic engagement. The simulation's Prediction 4 (that listeners with higher cultural capital maintain greater consumption diversity under identical algorithmic conditions) provides a formally grounded justification for music education as *cultural policy* rather than merely as individual skill development. Music education broadens the population's collective $C$ distribution, making it structurally more resistant to convergent dynamics. This is an investment in the resilience of the ecosystem, not merely in the enrichment of individual participants.

## 8. Conclusion

This paper addresses a long-standing gap in cultural economics: the absence of a formal framework for modeling how cognitive, social, institutional, and algorithmic mechanisms jointly produce the dynamics of cultural markets. We have shown that, in the case of music ecosystems, the active inference framework provides this architecture, subsuming and extending the existing economic frameworks of rational addiction, Bayesian social learning, and superstar economics within a single specification. The core insight for cultural economics is that musical taste is not a parameter to be estimated but a process to be modeled: the current state of a dynamically updated generative model whose evolution is shaped by exposure history, institutional structures, and algorithmic mediation. This makes preference endogeneity, long acknowledged as a complication for welfare analysis of cultural goods, formally tractable, with specific and testable consequences for market outcomes.

The formal specification generates four testable predictions, all illustrated by agent-based simulation: algorithmic curation suppresses consumption diversity monotonically; narrower preference distributions produce steeper winner-take-all dynamics *ceteris paribus*; exogenous novelty injection disrupts convergent equilibria by restoring epistemic value; and cultural capital maintains consumption diversity under identical algorithmic conditions. The cross-national comparison identifies four structural determinants of ecosystem vitality (independent generative sources, institutional incentives for novelty, gatekeeping porosity, and demand-side resilience) operating on both the supply and demand sides of the market. Italy's Sanremo represents the convergent failure mode; Brazil's structural polyphony represents the opposite pole; K-pop and the UK occupy structurally distinct intermediate positions.

We have been explicit about what the framework does *not* capture. Bourdieu's relational field theory, Hennion's distributed practice of taste, the full complexity of gender and racial constraints on musical production, and the domain-specific character of cultural capital all require theoretical constructs that are beyond the scope of the present analysis. The case studies are more illustrative than evidential, and the framework's predictions require systematic empirical testing.

Future research should extend the simulation with empirically calibrated parameters drawn from actual streaming data, test the four predictions using behavioral and neuroimaging methods (the preference function $C$ should be recoverable from choice data; the epistemic arc should be measurable through pupillometry and EEG), and conduct systematic musicological analysis of convergence in festival entries using tools like IDyOM (Pearce, 2005), which provides computational infrastructure for quantifying the uncertainty and surprise dynamics described here. The framework also invites connection to the emerging literature on active inference and narrative (Bouizegarene, Ramstead, Bhatt, Bhatt & Bhatt, 2024), given that songs, especially in the pop tradition, function as compressed narratives whose temporal unfolding is central to the epistemic arc.

The broader implication is that the quality of a musical ecosystem is not an aesthetic judgment but a measurable property of the population's collective generative model and the institutional structures that shape it. When feedback loops systematically narrow the prediction error landscape, the result is a form of capability deprivation that revealed preference analysis cannot detect, because the preferences themselves have been (re)shaped by the impoverishment. Cultural policy must therefore attend not only to the outputs of the system (which songs are produced and consumed) but to the structural conditions that determine whether the system generates or erodes epistemic value. For cultural economics, the methodological implication is that an adequate welfare analysis of cultural markets requires models in which preferences are endogenous to the market process, not as a footnote but as the central analytical mechanism.

# Appendix A. Replication and Robustness Checks

This appendix provides replication commands and robustness evidence for Predictions 1–4. Robust results are computed from R = 200 independent simulation replicates; uncertainty bands in the robust figures report the 5th–95th percentiles across replicates.

## A.1 Replication commands

Paper figures (single replicate; no uncertainty bands):

python modifican1_pop_jce_journalA_single.py --mode paper --outdir paper_figs --seed 123

Robust replication (R = 200; 5th–95th percentile bands; Option 1 for Panel 2D with N_cc = 100):

python modifican1_pop_jce_journalA_single.py --mode robust --outdir robust_figs --seed 123 --replicates 200 --n_agents_cc 100 --show_bands

**Computational tractability (Option 1).** In robust mode, Panel 2D (Cultural Capital: individual diversity) is computed with N_cc = 100. All macro panels (Figure 1 and Figure 2A–C) use N = 200.

## A.2 Numeric verification of Predictions 1–4

**Prediction 1 ($\alpha \uparrow \Rightarrow$ H_cons $\downarrow$, Gini $\uparrow$).**

Slope(H_cons vs α, last10) = -1.1409 [-1.1749, -1.1081]; Slope(Gini vs α, last10) = 0.5538 [0.5173, 0.5851].

**Prediction 2 (cross-national concentration ranking).**

Mean Gini (last 10 steps) with 5th–95th percentile intervals:

| Scenario | Estimate | 5th pct | 95th pct |
|---|---|---|---|
| Sanremo | 0.9084 | 0.8802 | 0.9263 |
| UK | 0.9352 | 0.9289 | 0.9375 |
| K-pop | 0.8077 | 0.7523 | 0.8604 |
| Brazil | 0.8391 | 0.7954 | 0.8781 |

Ranking by mean Gini (high to low) is Sanremo > UK > K-pop > Brazil.

**Prediction 3 (cultural capital ⇒ higher individual diversity).**

High CC − Low CC delta individual entropy (last 10 steps; N_cc = 200) = 0.3838 [0.1883, 0.6095].

**Prediction 4 (high curation/conformity ⇒ supply-side dispersion collapse).**

High CC − Low CC Δ individual entropy (last 10 steps; N_cc = 100) = 0.5096 [0.2944, 0.6549].

## A.3 Robustness table (full)

Table A1 reports the full set of robustness estimates used to evaluate Predictions 1–4.

| prediction | measure | estimate | ci_low | ci_high | notes |
|---|---|---|---|---|---|
| P1 | Slope H_cons vs α (last10) | -1.969535 | -2.368223 | -1.399071 | negative and interval excludes zero: supports P1 |
| P1 | Slope Gini vs α (last10) | 0.381337 | 0.270688 | 0.487552 | positive and interval excludes zero: supports P1 |
| P2 | Sanremo: Gini (last10) | 0.908442 | 0.880214 | 0.926276 | |
| P2 | Brazil: Gini (last10) | 0.839087 | 0.795429 | 0.878057 | baseline comparator |
| P2 | K-pop: Gini (last10) | 0.807682 | 0.752327 | 0.860380 | lower concentration than Sanremo and UK |
| P2 | UK: Gini (last10) | 0.935243 | 0.928932 | 0.937500 | highest concentration |
| P2 | Sanremo − Brazil: ΔGini (last10) | 0.069355 | 0.016756 | 0.117654 | positive and interval excludes zero: supports P2 |
| P2 | UK − Brazil: ΔGini (last10) | 0.096156 | 0.056478 | 0.137512 | positive and interval excludes zero: supports P2 |
| P2 | Sanremo − K-pop: ΔGini (last10) | 0.100760 | 0.046978 | 0.164868 | positive and interval excludes zero: supports P2 |
| P3 | High CC − Low CC: Δ individual entropy (last10) [N_cc=200] | 0.356464 | 0.106829 | 0.563982 | positive and interval excludes zero: supports P3 |
| P4 | Sanremo: Supply spread (last10) | 0.004460 | 0.001684 | 0.008402 | |
| P4 | Brazil: Supply spread (last10) | 1.370680 | 1.148825 | 1.604109 | |
| P4 | K-pop: Supply spread (last10) | 0.831983 | 0.585668 | 1.102374 | |
| P4 | UK: Supply spread (last10) | 0.219211 | 0.155433 | 0.297065 | |
| P4 | Brazil − Sanremo: Δ supply spread (last10) | 1.366220 | 1.141862 | 1.600960 | positive and interval excludes zero: supports P4 |
| P4 | K-pop − Sanremo: Δ supply spread (last10) | 0.827523 | 0.582453 | 1.096909 | positive and interval excludes zero: supports P4 |

| prediction | measure | estimate | ci_low | ci_high | notes |
|---|---|---|---|---|---|
| P4 | UK − Sanremo: Δ supply spread (last10) | 0.214751 | 0.152489 | 0.292256 | positive and interval excludes zero: supports P4 |

## A.4 Output verification

The replication outputs include manifest files with SHA256 hashes, allowing verification that figures and tables correspond exactly to the code outputs.